# THE HIGH FIELD MAGNETIZATION IN THE RF SPUTTER DEPOSITED COPPER FERRITE THIN FILMS

Prasanna D. Kulkarni and Shiva Prasad
Department of Physics
IIT Bombay, Mumbai 400076

N. Venkataramani
ACRE IIT Bombay
Powai, Mumbai 400076

R. Krishnan
Laboratoire de Magnetisme et d’optique de Versailles,CNRS, 78935 Versailles, France

Wenjie Pang, Ayon Guha, R.C. Woodward, R.L. Stamps
School of Physics, M013, The University of Western Australia, 35 Stirling Hwy, Crawley WA 6009 (Australia)

## ABSTRACT

Copper ferrite thin films were deposited on amorphous quartz substrates. The as deposited films were annealed in air and either quenched or slow cooled. Magnetization studies were carried out on the as deposited as well as annealed films using a SQUID magnetometer. The M-H curves were measured up to a field of 7T, at temperatures varying from 5K to 300K. The magnetization in the films did not saturate, even at the highest field. The expression, $M(H)= Q(1- a /H^n)$ fitted the approach to saturation best with n=1/2, for all films and at all temperatures. The coefficient *a* was the highest for the as deposited film and was the smallest for the quenched film. In the case of as deposited film, the value of coefficient *a* increased with increasing temperature, while for the annealed films, the value of *a* showed a decrease as temperature increases.

## INTRODUCTION

The magnetization of nanocrystalline ferrite films does not saturate even at very high fields. This phenomenon is called the High Field Susceptibility (HFS). Margulies et al. have reported non-saturation in magnetic fields upto 8T in $Fe_3O_4$ samples[1], at room temperature. A detailed study of HFS in LiZn ferrite films at room temperature is recently published by Dash et al.[2]. They attribute this phenomenon to the defects in the thin films, even though their exact origin and nature are not discussed. In order to understand the reason for this phenomenon, it is necessary to get more data on this behavior.

In the present paper, a study of the temperature dependence of the HFS on copper ferrite thin films is reported. Copper ferrite can be stabilized in two different phases in thin film form, viz., a cubic and a tetragonal phase[3]. Quenching the copper ferrite films after carrying out a post deposition annealing stabilizes the cubic phase. The slow cooling of the films after the annealing, on the other hand, results in the tetragonal phase. High field magnetization studies are carried out on both cubic and tetragonal phases of the copper ferrite films.

**EXPERIMENTAL**

Copper ferrite films were deposited using a Leybold Z400 rf sputtering system on amorphous quartz substrates. No heating or cooling was carried out during sputtering. The rf power employed during the deposition was 50W. The thickness of the films was ~2400 Å. The films were also annealed at 800°C for 2 hours, followed by either quenching or slow cooling. The magnetization was measured using a SQUID magnetometer for as deposited, quenched and slow cooled films in a field up to 7T and at various temperatures between 5K and 300K.

**RESULTS**

Fig.1(a), shows the magnetization M(H) as a function of field H for the as deposited (Asd), slow cooled (SC) and quenched (Que) film at 300K. One can notice from the figure that the magnetization of the Que film is the highest. This is because of the cubic nature of the film[3]. One also notices clearly that the magnetization of the films does not saturate even at the highest field for all the three films.

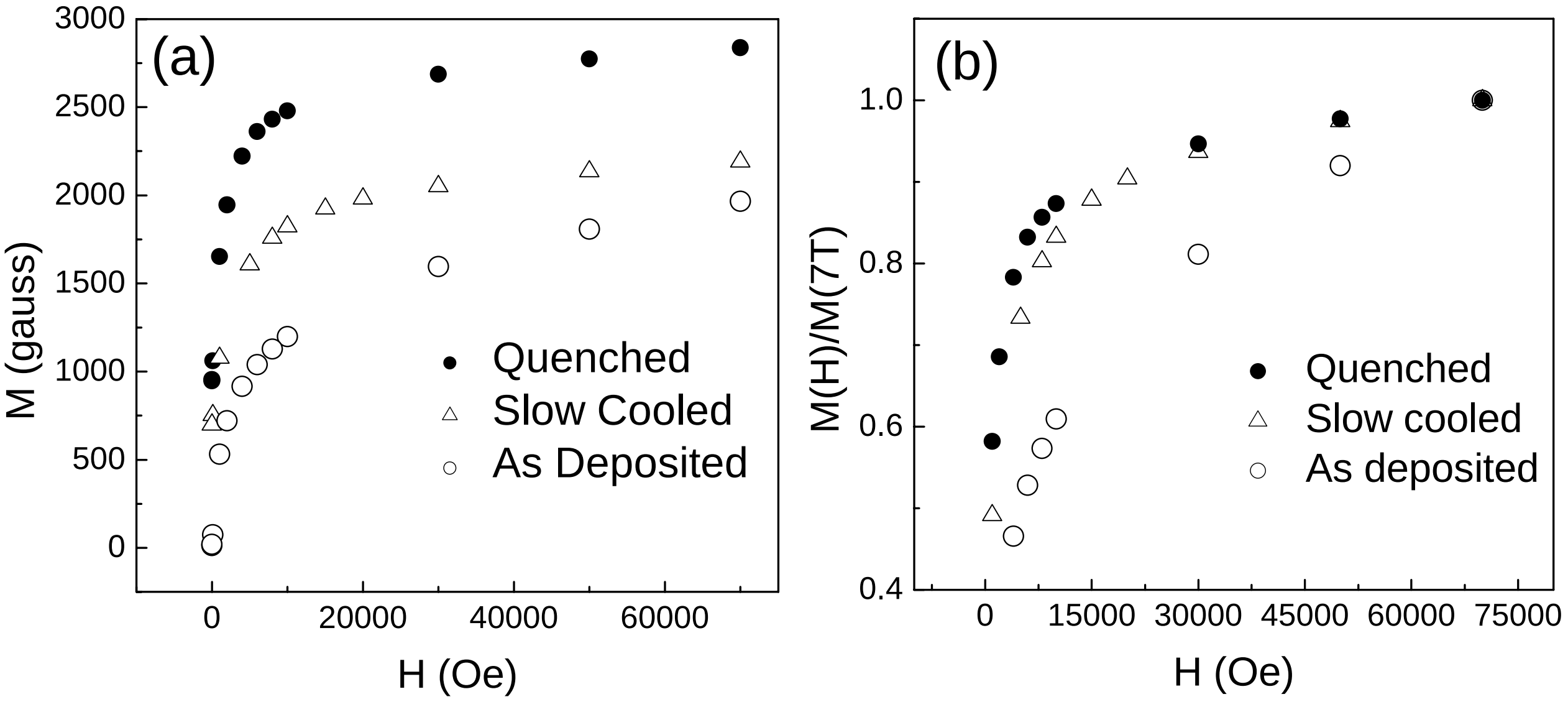

Fig. 1. (a) M Vs H, (b) M(H)/M(7T) Vs H, for 50W copper ferrite thin films at 300K.

To assist the comparison of HFS, the data of Fig.1(a) is replotted in Fig.1(b), with M(H)/M(7T) on the y-axis. One can very clearly see from these figures that the Asd film has the least tendency to saturate, while the Que film seems to get saturated comparatively easily.

In order to see the effect of temperature, M(H)/M(7T) has been plotted in Fig. 2 (a) (b) and (c), as a function of field at 5K and 300K for the Asd, SC and Que films. From Fig. 2, one observes an interesting feature related to the temperature dependence of high field susceptibility. While the HFS increases upon lowering the temperature in SC and Que films, it decreases for the Asd film.

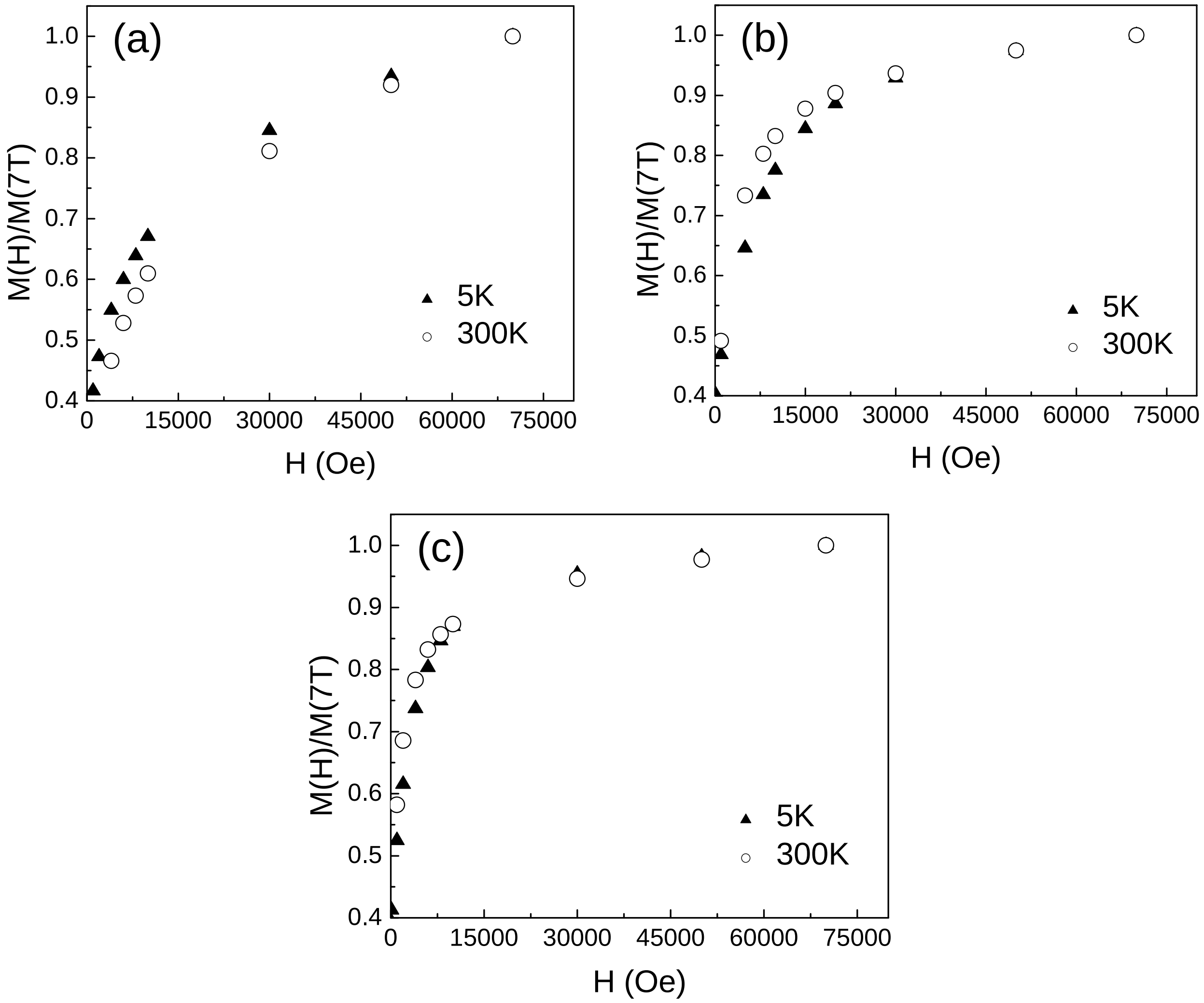

Fig. 2. M(H)/M(7T) as a function of magnetic field at 5K and 300K for (a) As deposited, (b) Slow cooled, (c) Quenched, 50W copper ferrite thin films.

DISCUSSION

The approch to saturation is discussed by Chikazumi[4]. In the Chikazumi expression, the magnetization at a given applied field is written in the following way,

$$4\pi M = Q\,(1 - a/H^{1/2} - b/H - c/H^2 - \ldots.) + eH \qquad (1)$$

Here $4\pi M$ is the actual value of magnetization that is observed at a field H, and Q, a, b, c, and e are constants. The value of Q should correspond to a value of $4\pi M$ in the infinite field. The last term, eH, is caused by an increase in spontaneous magnetization by the external field and is generally assumed to be negligible. The $c/H^2$ term is related to the presence of restoring torque due to the magnetic anisotropy. Chikazumi has derived the expression for the coefficient *c* for the case when the torque due to external field counterbalances the restoring torque due to magnetic anisotropy. This coefficient *c* is related to the anisotropy constant and magnetization of the material. The observed b/H term can be explained if the restoring torque increases with the approach of magnetization to saturation. If the magnetization is fixed firmly due to point defects, the magnetization surrounding these defects will form transition layers that are similar to the ordinary domain wall. The thickness of these transition layers will decrease with the increase in field strength. In this case, the change of magnetization will be first proportional to the $1/H^{1/2}$ term and finally to the $1/H^2$ term. Dash et al[2] used an expression of the type $a/H^{1/2}$ to fit their magnetization data at high field and found that it fits their data best up to the highest field. Based on that they concluded that high field magnetization in their films is caused by point defects.
In order to quantify the high field susceptibility we have fitted the magnetization data from 0.8T to 7T field to an expression involving just the $a/H^{1/2}$ term in eq. (1). The value of *a* thus found is plotted in Fig. 3 as a function of temperature for Asd, SC and Que films. One can see that the value of *a* is the smallest for the Que films, which are cubic.

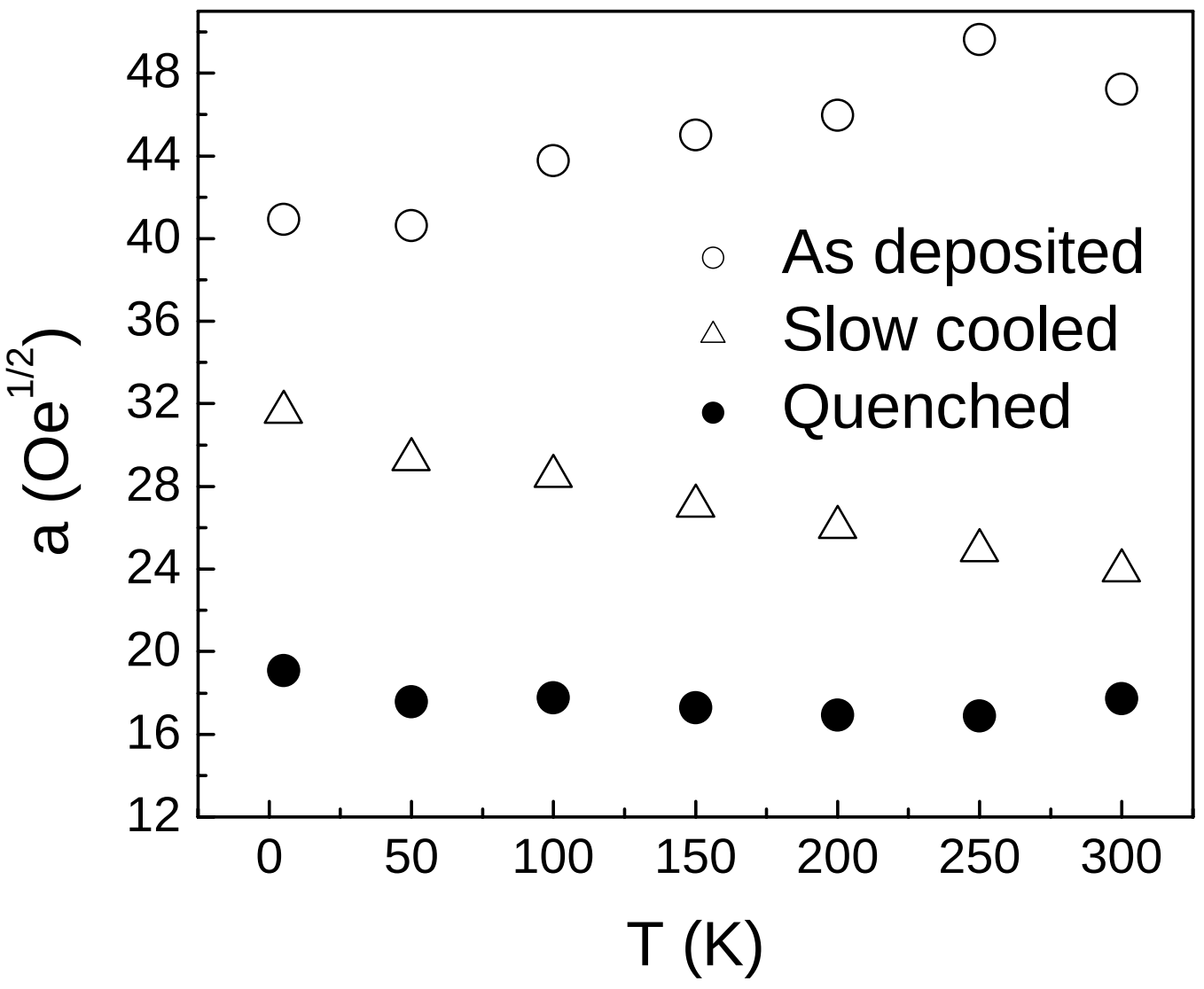

Fig. 3. The variation of the coefficient *a* of $H^{1/2}$ term with temperature for 50W films.

If the HFS phenomenon is due to the presence of defects as envisaged by Chikazumi, one indeed expects that the slope *a* would be largest in Asd film, as these films are likely to be most defective. Annealing is expected to cause reduction in the defects, which can explain why the slope decreases in the case of Que and SC film. Even in annealed films, the value of *a* is smaller for Que film than SC film. The Que film being cubic has lesser anisotropic field than the SC film, which is in tetragonal phase. This indicates *a* could be related to magnetocrystalline anisotropy. If it is so *a* is also expected to increase with decreasing temperature which is observed.

It is known that the sputter deposited ferrite films are nano-crystalline[5]. The grain sizes in the films are so small that an isolated grain of that size would be superparamagnetic. However a M-H loop is observed in all these materials indicating the presence of magnetic order. This is most likely because of the inter-granular interactions, which may substantially modify the superparamagnetic behavior. Nevertheless the film may contain grains, which are superparamagnetic in addition to the one that are magnetically ordered. These superparamagnetic grains may also contribute to the HFS.

If HFS is due to superparamagnetism, then also one can understand the lower value of *a* in annealed films. The fraction of the superparamagnetic particles is likely to be less in the SC and Que film than in the Asd film. This is due to the increase in the grain sizes of the material with annealing. The annealed films still remain nano-crystalline[5]. However, in the case of superparamagnetic particles, the susceptibility increases with the decrease of temperature following Curie's law[6].
If superparamagnetism is dominant in Asd films, *a* is expected to increase with lowering temperature. The fact that this is not observed could be either because some grains get frozen into magnetic order upon lowering the temperature or because of some effect involving intergranular interaction. The value of the coefficient *a* for the Que and the SC films, on the other hand, shows a behavior, which is somewhat similar to the susceptibility of superparamagnetic particles. But these are the films in which the grain sizes are expected to be bigger, and they are likely to contain lesser number of superparamagnetic grains. The present results, thus, can not be understood purely in terms of superparamagnetic grains. A combination of superparamagnetic and anisotropic effects may be required to explain the HFS phenomenon.

## CONCLUSIONS

The coefficient of $a/H^{1/2}$ term has the highest value for as deposited film and is lowest for quenched film. The temperature dependence of *a* shows that the value of a decreases with decreasing temperature for as deposited film. For annealed films, however, *a* increases with decreasing temperature. This can not be understood purely on the basis of presence of superparamagnetic grains. A combination of defects and superparamagnetism could be responsible.

## ACKNOWLEDGEMENT

The author Prasanna D. Kulkarni thanks the CSIR, India for financial support.

## REFERENCES

[1]D.T. Margulies, F.T. Parker, F.E. Spada, R.S. Goldman, J. Li, R. Sinclair, A.E. Borkowitz, "Anomalous moment and anisotropy behavior in $Fe_3O_4$ flims," *Physical Review B,* 53[14] 9175-9187 (1996-II).

[2]J. Dash, S. Prasad, N. Venkataramani, R. Krishnan, P. Kishan, N. Kumar, S.D. Kulkarni, S.K. Date, "Study of magnetization and crystallization in sputter deposited LiZn ferrite thin films," *Journal of Applied Physics*, 86[6] 3303-3311 (1999).

[3]M.Desai, S. Prasad, N. Venkataramani, I. Samajdar, A.K. Nigam, R. Krishnan, "Annealing induced structural change in sputter deposited copper ferrite thin films and its impact on magnetic properties," *Journal of Applied Physics*, 91[4] 2220-27 (2002).

[4]S. Chikazumi, S.Charap, "Law of approach to saturation,"; pp. 274-279 in *Physics of Magnetism*, John Wiley & Sons, New York (1964).

[5]M. Desai, J. Dash, I. Samajdar, N. Venkataramani, S. Prasad, P. Kishan, N. Kumar, "A TEM study on lithium zinc ferrite thin films and the microstructure correlation with the magnetic properties," *Journal of Magnetism and Magnetic Materials,* 231 108-112 (2001).

[6]C. P. Bean, J.D. Livingston, " Superparamagnetism," *Journal of Applied Physics,* 30[4] 120S-129 (1959).